\documentclass[aps,prl,twocolumn,psfig,longbibliography,superscriptaddress]{revtex4}

\usepackage[compat=1.0.0]{tikz-feynman}
\usepackage{amssymb}
\usepackage{multirow}
\usepackage{bm}
\usepackage{amsmath}
\usepackage{graphicx}
\usepackage{epstopdf}
\usepackage{subfigure}
\usepackage{natbib}
\usepackage{epsfig}
\usepackage{amsfonts}
\usepackage{mathrsfs}
\usepackage{braket}
\usepackage[toc,page,title,titletoc,header]{appendix}
\usepackage[colorlinks,linkcolor=blue,citecolor=blue,anchorcolor=blue]{hyperref}
\usepackage{dsfont,amsthm,amsbsy}

\usepackage{fancyhdr}

\usetikzlibrary{quotes,angles}
\newcommand{\obtuseangle}{\kern.08em
\begin{tikzpicture}
    \draw coordinate (a) at (0.14,0);
    \draw coordinate (b) at (0,0);
    \draw coordinate (c) at (-.12,0.18);
    \draw (a) -- (b) -- (c) pic [draw=black]{} ;
\end{tikzpicture}%
\kern.08em%
}

\begin{document}
\title{Pair density wave and superconductivity in a kinetically frustrated doped Emery model on a square lattice}
\author{Hong-Chen Jiang}
\email{hcjiang@stanford.edu}
\affiliation{Stanford Institute for Materials and Energy Sciences, SLAC National Accelerator Laboratory and Stanford University, Menlo Park, California 94025, USA}
\author{Thomas Peter Devereaux}
\email{tpd@stanford.edu}
\affiliation{Stanford Institute for Materials and Energy Sciences, SLAC National Accelerator Laboratory and Stanford University, Menlo Park, California 94025, USA}
\affiliation{Dept. of Materials Science and Engineering, Stanford University, Stanford, California 94305, USA}

\date{\today}
\begin{abstract}
The quest to understand the nature of superconductivity in cuprates has spotlighted the pair density wave (PDW) -- a superconducting state characterized by a spatially modulated order parameter. Despite significant advances in understanding PDW properties, conclusively demonstrating its presence in systems pertinent to cuprate superconductors remains elusive. In this study, we present a systematic density-matrix renormalization group study to investigate the Emery model (or the three-band Hubbard model) on two-leg square cylinders with negative electron hopping term $t_{pp}$ between adjacent oxygen sites. Kinetic frustration - introduced by changing the sign of oxygen-oxygen hopping - leads to a much reduced Cu-Cu antiferromagnetic exchange along with an enlarged charge transfer energy that changes the local properties of the model. At light doping levels, our findings reveal a ground state remarkably consistent with a PDW, exhibiting mutually commensurate superconducting (SC), charge and spin density wave correlations. Intriguingly, the dominant SC pairing is observed between neighboring oxygen sites, diverging from the expected Cu sites in the positive $t_{pp}$ case. When the system incorporates moderate near-neighbor interactions, particularly an attractive $V_{pp}$ between adjacent oxygen sites, the SC correlations become quasi-long-ranged, accompanied by a pronounced divergence in the PDW susceptibility. When further increase the attractive $V_{pp}$, the system gives ways to an unconventional $d$-wave superconductivity.
\end{abstract}
\maketitle

\begin{figure}
  \includegraphics[width=1.0\linewidth]{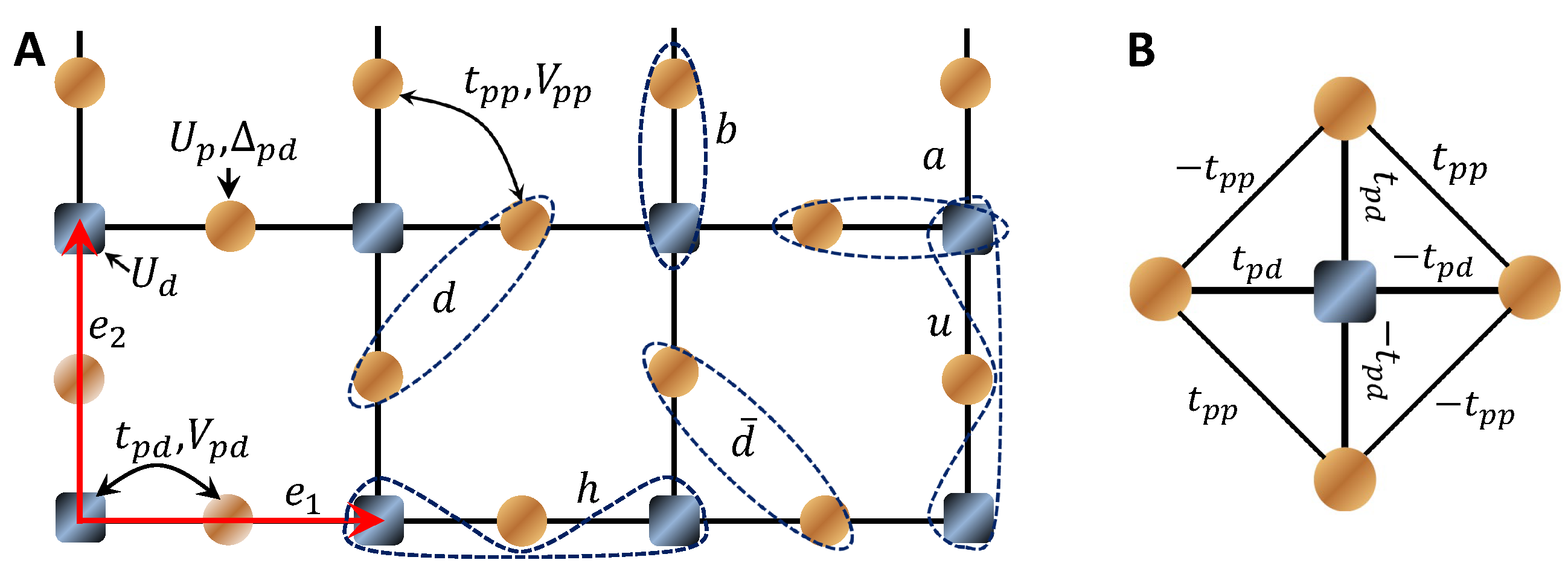}
  \caption{Emery model on the square lattice. (A) The squares represent Cu $d_{x^2-y^2}$ orbitals and circles represent O$_{x/y}$ 2$p_x$/2$p_y$ orbitals. Periodic (open) boundary condition is imposed in the direction specified by the lattice basis vector $e_2=(0,1)$ ($e_1=(1,0)$). The dashed loops represent bonds $a$, $b$, $d$, $\bar{d}$, $h$ and $u$. (B) Signs of hopping matrix elements in an elementary plaquette in the CuO$_2$ plane.} \label{Fig:Lattice}
\end{figure}

\textbf{Introduction -- }
The Emery model, also knows as the three-band Hubbard model, has long been proposed to as one of the minimal models to understand the electronic properties of cuprate high-temperature superconductors \cite{Zaanen1985,Emery1987,Scalettar1989,Scalettar1991,White2015,Huang2017,Jiang2023DF}. In this model, a square lattice of copper (Cu) and oxygen (O) atoms in the CuO$_2$ plane (see Fig.\ref{Fig:Lattice}) is considered, where the Copper sites are represented by a single $3d_{x^2-y^2}$ orbital, while each oxygen site has two active $2p$ orbitals ($2p_x$ and $2p_y$). In the hole representation, the model Hamiltonian is defined as%
\begin{eqnarray}
H &=& H_k+\Delta_{pd}\sum_{i\sigma}\hat{p}_{i\sigma}^+\hat{p}_{i\sigma}+U_d\sum_i \hat{n}_{i\uparrow}^d \hat{n}_{i\downarrow}^d \nonumber\\%
&+&U_p\sum_i \hat{n}_{i\uparrow}^p \hat{n}_{i\downarrow}^p  +V_{pd}\sum_{\langle ij\rangle}\hat{n}_i^d \hat{n}_j^p + V_{pp}\sum_{\langle ij\rangle}\hat{n}_i^p \hat{n}_j^p.\\%
H_k &=& \sum_{\langle ij\rangle\sigma}t_{pd}^{ij}(\hat{d}_{i\sigma}^+\hat{p}_{j\sigma}+h.c.)+\sum_{\langle ij\rangle\sigma}t_{pp}^{ij}(\hat{p}_{i\sigma}^+\hat{p}_{j\sigma}+h.c.)\nonumber%
\label{Eq:Ham}
\end{eqnarray}
Here $\hat{d}_{i\sigma}^+$ and $\hat{p}_{j\sigma}^+$ create holes with spin-$\sigma$ on the $i^{th}$ Cu and $j^{th}$ oxygen sites, and $\langle ij\rangle$ denotes NN sites. $\hat{n}_{i\sigma}^d=\hat{d}_{i\sigma}^+ \hat{d}_{i\sigma}$ and $\hat{n}_{i\sigma}^p=\hat{p}_{i\sigma}^+ \hat{p}_{i\sigma}$ are the number operators for spin-$\sigma$ at the Cu and O sites, respectively, with the total number operators are defined as $\hat{n}_i^{d}=\sum_\sigma \hat{n}_{i\sigma}^{d}$ and $\hat{n}_i^{p}=\sum_\sigma \hat{n}_{i\sigma}^{p}$. $\Delta_{pd}$ is the energy difference between having a hole on the Cu and oxygen sites. $t_{pd}^{ij}$ and $t_{pp}^{ij}$ are the hole hopping matrix elements between nearest-neighbor (NN) Cu and oxygen sites and the NN oxygen sites, respectively. $U_d$ and $U_p$ are the on-site Cu and oxygen Coulomb repulsion, and $V_{pd}$ and $V_{pp}$ are the NN Cu-O and O-O Coulomb interactions, respectively.

While the Emery model has been proposed as one of the critical framework for studying the cuprates superconductors, which captures phenomena like superconductivity, charge and spin density wave orders \cite{Zaanen1985,Emery1987,Scalettar1989,White2015,Huang2017,Jiang2023DF}, it has more recently been extended to investigate the emergence of novel pair density wave (PDW) states \cite{Jiang2023PDW}. In a PDW state, the superconducting (SC) order parameter carries finite center-of-mass momentum and varies spatially so that its spatial average vanishes \cite{FF1964,LO1965,Berg2009,Fradkin2015,Agterberg2020,Lee2014,Jian2020,Lozano2022}. The PDW state has been considered as a promising candidate state to understand the physics of cuprates high-temperature superconductors and other strongly correlated systems, where it has been proposed that various phases, including the superconductivity, charge and spin density wave orders, can emerge by partially melting the PDW state \cite{Lee2014,Fradkin2015,Agterberg2020,Himeda2002}. Recently, intense interest in the PDW state has emerged due to experimental observations in cuprate superconductors ${\rm Bi_2Sr_2CaCu_2O_{8+x}}$ \cite{Hamidian2016,Ruan2018,Edkins2019,Liu2021} and La$_{1.875}$Ba$_{0.125}$CuO$_4$ \cite{Agterberg2008,Li2007,Berg2007,Tranquada2008,Tranquada2020,Tranquada2021}, kagome superconductor CsV$_3$Sb$_5$ \cite{Chen2021k} and ion-based superconductor \cite{Liu2023}.

The realization of PDW state in microscopic lattice models remain highly nontrivial and usually involves modifying or extending existing frameworks like the Hubbard models to include competing interactions or inhomogeneities that can give rise to spatial modulation \cite{Berg2010,Fradkin2012,Venderley2019,Xu2019,Peng2021a,Peng2021b,Han2020,Huang2022,Wu2023,Jiang2023YF}. These include the Kondo-Heisenberg model \cite{Berg2010}, the extended Hubbard-Heisenberg model \cite{Fradkin2012}, the strong coupling limit of the Holstein-Hubbard model \cite{Han2020,Huang2022} and generalized $t$-$J$ and Hubbard models \cite{Xu2019,Venderley2019,Peng2021a,Peng2021b}. More recently, it has also been shown by one of us that the PDW ground state can also be realized in the three-band Hubbard model on a two-leg square cylinder \cite{Jiang2023PDW}, where the SC correlations are dominant between neighboring Cu sites with $d_{x^2-y^2}$-wave pairing symmetry.

\textbf{Model and method -- }%
In the present work, we consider the Emery model on the square lattice as defined in Fig.\ref{Fig:Lattice} and Eq.(\ref{Eq:Ham}) to study whether the same PDW state or distinct SC state emerges upon doping as well as the associated pairing symmetry using the density-matrix renormalization group (DMRG)\cite{White1992,White1993}. The signs of the hopping matrix elements in the related orbital configuration, i.e., Cu $3d_{x^2-y^2}$ orbital and O$_{x/y}$ $2p_x/2p_y$ orbitals, of an elementary plaquette centered at a generic Cu site is shown in Fig.\ref{Fig:Lattice}B. It is noted that the sign of $t_{pp}\equiv t_{pp}^\sigma-t_{pp}^\pi$ is taken to be negative, opposite to that is usually chosen \cite{Eskes1990}. The resulting increase in the delocalization energy involving ligand $L$ oxygen orbitals raises the level of the effective charge transfer energy, alters the magnetic exchange among Cu and O, and affects the local symmetry of the ground state orbital configuration. This has a profound impact on the local physics and ground state properties of the system. 

Following Ref. \cite{White2015,Jiang2023PDW}, we set $t_{pd}=1$ as the energy unit and take a canonical set of parameters $U_d=8$, $U_p=3$, $\Delta_{pd}=3$ for cuprates \cite{White2015,Armitage2010,Haule2014} but negative $t_{pp}=-0.5$, and study the ground state properties of this system as a function of $V_{pd}$ and $V_{pp}$. We focus on two-leg cylinders as shown in Fig.\ref{Fig:Lattice} with width $L_y=2$ and length up to $L_x$=96, where $L_x$ and $L_y$ are the number of unit cells along the $e_1$ and $e_2$ directions, respectively. The total number of sites is $N=3L_x L_y+2L_y=3N_u+2L_y$, where $N_u$ is the number of unit cell. The overall hole density of the system is defined as $\rho=1+\delta$, where $\delta=N_h/N_u$ and $N_h$ denote the hole doping concentration and number of doped holes away from half-filling, respectively. We consider $\delta=1/12$ and $1/8$, and keep up to $m=20000$ states with a typical truncation error $\epsilon\sim 10^{-10}$.

\begin{figure}[!th]
\centering
  \includegraphics[width=1\linewidth]{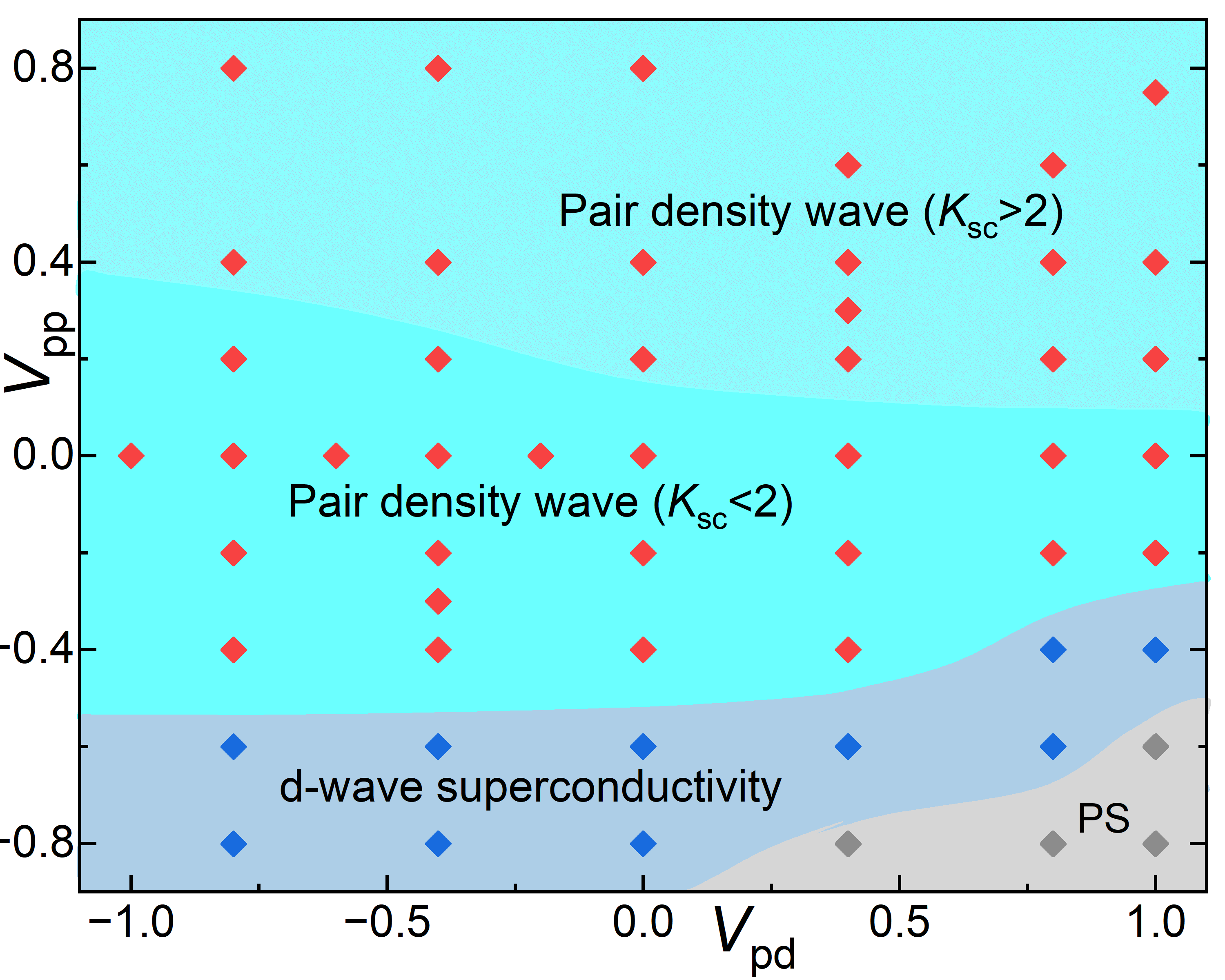}
\caption{Ground state phase diagram of the Emery model on two-leg square cylinders at $\delta=1/8$. The solid symbols are numerical data points and PS denotes phase separation. The shaded regions are guides for eyes.}\label{Fig:Phase}
\end{figure}

\textbf{Phase diagram -- }%
Our main results are summarized in the ground state phase diagram in Fig. \ref{Fig:Phase}. When $V_{pp}$ between oxygen sites are not strongly attractive, we find the the ground state of the system is consistent with that of a PDW state with power-law and mutually commensurate SC, charge-density wave (CDW) and spin-density-wave (SDW) correlations. The SC correlations oscillate periodically in real space in such a way that its spatial average vanishes and the PDW ordering wavevector $Q\approx 2\pi\delta$ is incommensurate. Contrary to the positive $t_{pp}$ case \cite{White2015,Jiang2023PDW}, our results show that the SC pairing is dominant between adjacent oxygen sites instead of Cu sites. Accordingly, the SC pairing symmetry is consistent with $d_{xy}$ rather than $d_{x^2-y^2}$ wave. Similar to the single-band Hubbard model on the square lattice \cite{Chen2021,Qu2022,Peng2023}, the finite electronic attractions $V_{pd}$ and $V_{pp}$, especially $V_{pp}$ between oxygen sites, can notably enhance the SC correlations while simultaneously suppress the CDW correlations. For modestly strong $V_{pp}$ interaction, including both repulsion and attraction, the SC correlations become strong enough so that a quasi-long-range PDW order emerges with $K_{sc}<2$ and divergent static PDW susceptibility. When further increase the attractive $V_{pp}$, the system gives ways to the $d$-wave superconductivity where the PDW signatures get much suppressed. Interestingly, similar with the PDW phase, the Cooper pairing between the adjacent oxygen sites also dominates in the pairing channel and the corresponding pairing symmetry is consistent with the $d_{xy}$-wave as well. For even stronger attractive $V_{pp}$, the ground state of the system becomes phase separated.

\begin{figure}[!th]
\centering
  \includegraphics[width=1\linewidth]{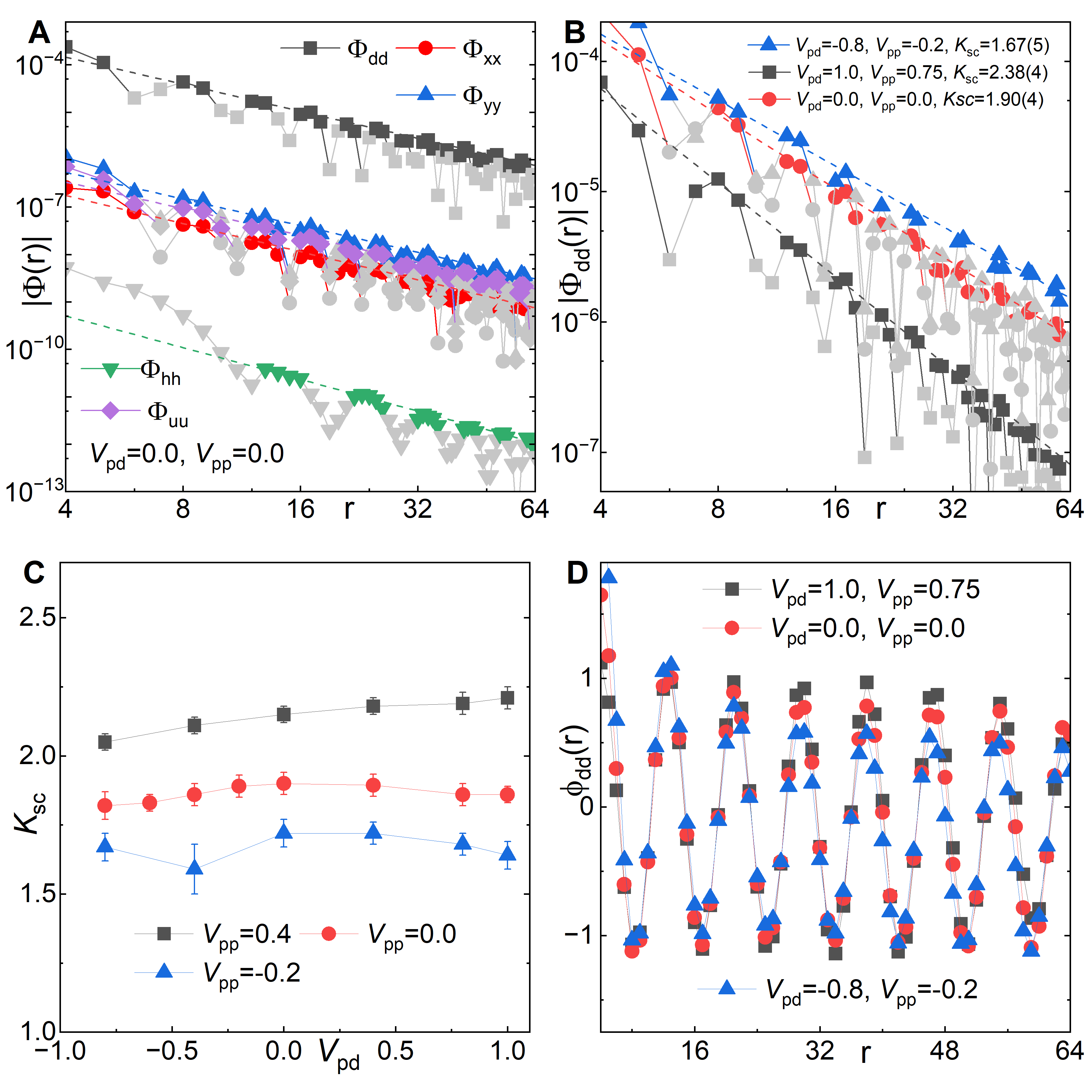}
    \caption{Superconducting correlations in the pair density wave phase at $\delta=1/8$. The magnitude of SC correlations are shown in (A) for $\Phi(r)$ at $V_{pd}=V_{pp}=0$, and in (B) for $\Phi_{dd}(r)$ with different $V_{pd}$ and $V_{pp}$. The dashed lines represent fits to a power-law function $f(r)\sim r^{-K_{sc}}$. Data points far from the envelope and those at short distances are discarded in gray color in the fitting process. (C) Luttinger exponent $K_{sc}$ as a function of $V_{pd}$ and $V_{pp}$. (D) The normalized functions $\phi_{dd}(r)=\Phi_{dd}(r)/f_{dd}(r)$ reflect the spatial oscillation of $\Phi_{dd}(r)$.}\label{Fig:PDW}
\end{figure}

\begin{figure}
  \includegraphics[width=1\linewidth]{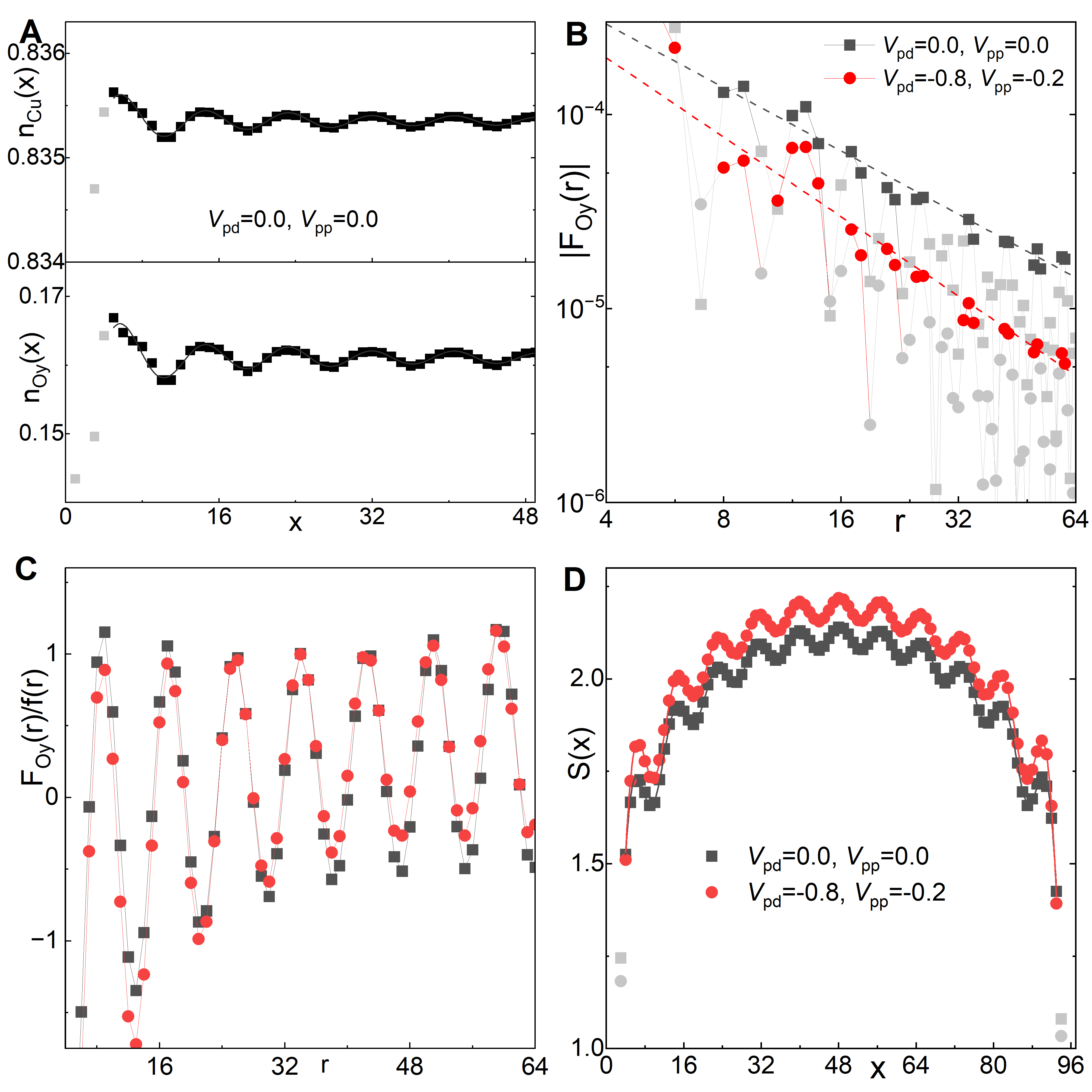}
  \caption{Charge density profile, spin-spin correlation and entanglement entropy in the pair density wave phase at $\delta=1/8$. (A) Charge density profiles $n_{Cu}(x)$ on the Cu site and $n_{Oy}(x)$ on the Oy site for $V_{pd}=V_{pp}=0$. (B) The magnitude of the spin-spin correlation $|F(r)|$ where the dashed lines represent power-law fits $f(r)\sim r^{-K_s}$. (C) The normalized function $F(r)/f(r)$ reflects the spatial oscillation of $F(r)$ in (B). (D) Von Neumann entanglement entropy $S(x)$. Note that a few data points in gray color close to the open ends are excluded to minimize boundary effects.}\label{Fig:NxSS}
\end{figure}

\textbf{Pair density wave phase -- }%
As shown in Fig. \ref{Fig:Phase}, the majority of the ground state phase diagram is occupied by the PDW phase,where the SC correlations decay as a power-law at long distances and oscillate periodically in real space in such a way that its spatial average vanishes. We provide detailed examples in Fig. \ref{Fig:PDW} and Fig. \ref{Fig:NxSS} for several characteristic sets of parameters. Our conclusions hold for all parameters in the PDW phase in Fig. \ref{Fig:Phase}.

\textit{Superconducting correlations -- }%
In order to explore the potential for superconductivity, we have calculated the equal-time spin-singlet SC pair-pair correlations defined as%
\begin{eqnarray}\label{Eq:SC_cor}
\Phi_{\alpha\beta}(r)=\langle\hat{\Delta}^{\dagger}_{\alpha}(x_0,y_0)\hat{\Delta}_{\beta}(x_0+r,y_0)\rangle.
\end{eqnarray}
Here, $\hat{\Delta}^{\dagger}_{\alpha}(x,y)=\frac{1}{\sqrt{2}}[\hat{c}^{\dagger}_{(x,y),\uparrow}\hat{c}^{\dagger}_{(x,y)+\alpha,\downarrow}-\hat{c}^{\dagger}_{(x,y),\downarrow}\hat{c}^{\dagger}_{(x,y)+\alpha,\uparrow}]$ is spin-singlet pair creation operator on the bond $\alpha=a$, $b$, $d$, $\bar{d}$, $h$ and $u$ defined in Fig.\ref{Fig:Lattice}A. ($x_0,y_0$) is a reference bond with $x_0\sim L_x/4$, $r$ is the distance between two bonds in the $e_1$ direction. We have comprehensively analyzed the various components of the SC correlations. This includes calculations of $\Phi_{aa}$, $\Phi_{ab}$, $\Phi_{bb}$, $\Phi_{dd}(r)$, $\Phi_{d\bar{d}}(r)$, $\Phi_{\bar{d}\bar{d}}(r)$, $\Phi_{hh}$, $\Phi_{uu}$ and $\Phi_{uh}$. While the positive $t_{pp}$ case primarily showcases dominant correlations in $\Phi_{hh}$ and $\Phi_{uu}$ as discussed in \cite{Jiang2023PDW}, our results differ significantly for the negative $t_{pp}$ case. As illustrated in Fig. \ref{Fig:PDW}A, we observe that the strongest SC correlations are prominently exhibited in $\Phi_{dd}(r)$ and $\Phi_{\bar{d}\bar{d}}(r)$. This suggests that the pairing is more dominant between neighboring oxygen sites, rather than Cu sites. Furthermore, even though the pairing symmetry aligns with the $d$-wave, our findings indicate a shift to the $d_{xy}$-wave symmetry in this particular scenario, diverging from the previously understood $d_{x^2-y^2}$-wave. This distinction in $d_{xy}$-wave symmetry is characterized by the relationship: $\Phi_{dd}(r) \sim \Phi_{\bar{d}\bar{d}}(r) \sim -\Phi_{d\bar{d}}(r)$."

We've closely examined the spatial distribution of SC correlations, specifically targeting $\Phi_{dd}(r)$. Our findings, based on three representative parameter choices, are depicted in Fig. \ref{Fig:PDW}D. Here, $\Phi_{dd}(r)$ exhibits clear spatial oscillations as $\Phi_{dd}(r) \sim f(r)\phi_{dd}(r)$ over a vast region of $r$. In this context, $f(r)$ acts as the envelope, while $\phi_{dd}(r)$ gives rise to the spatial oscillation. As we move to longer distances, the envelope function $f(r)$ adheres to a power-law decay $f(r) = A \ast r^{-K_{sc}}$ \cite{Kuhner2000}. For instance, we derived an exponent $K_{sc}\approx 2.4$ at $V_{pd}=1.0$ and $V_{pp}=0.75$, and $K_{sc}\approx 1.9$ at $V_{pd}=0$ and $V_{pp}=0$, and $K_{sc}\approx 1.7$ at $V_{pd}=-0.8$ and $V_{pp}=-0.2$. Drawing connections with the established single-band Hubbard model \cite{Chen2021,Qu2022,Peng2023} and the positive $t_{pp}$ Emery model \cite{Jiang2023PDW}, we find that diminishing the NN repulsion or amplifying the NN attraction, especially $V_{pp}$, can notably enhance SC correlations. This observation is further validated by the $K_{sc}$ values presented in Fig. \ref{Fig:PDW}C. More comprehensive results of $K_{sc}$ for $\Phi_{dd}$ at $\delta=1/8$ are shown in Fig. \ref{Fig:PDW}C. These point towards the divergence of the static PDW susceptibility, characterized as $\chi_{pdw} \sim T^{-(2-K_{sc})}$ when $T \rightarrow 0$. We have also calculated the spin-triplet SC correlations, which are markedly weaker than their spin-singlet counterparts.

The spatial oscillation of the SC correlations $\Phi(r)$ is captured by the normalized function $\phi(r)$, as previously defined. Depictions of $\phi_{dd}(r)$ are presented in Fig. \ref{Fig:PDW}D and align well with the fitting function $\phi_{dd}(r) \sim \sin(Qr + \theta)$. This pattern resonates with characteristics observed in the PDW state with vanishing spatial average of $\phi(r)$ \cite{Agterberg2020}. The PDW ordering wavevector appears to be incommensurate as $Q \approx 2\pi\delta$ with a corresponding wavelength $\lambda_{sc} \approx 1/\delta$. For instance, $\lambda_{sc}\approx 8$ for $\delta=1/8$ as evidenced in Fig. \ref{Fig:PDW}D, whereas $\lambda_{sc}\approx 12$ for $\delta=1/12$.

\textit{Charge density wave -- }%
We have calculated the charge density profile $n_\alpha(x,y)=\langle \hat{n}_\alpha(x,y)\rangle$ and its rung average $n(x)={\sum_{y=1}^{L_y}} n_a(x,y)/L_y$ (e.g., Fig. \ref{Fig:NxSS}A) to describe the charge density properties of the system, where $\alpha$=Cu/O$_x$/O$_y$ site. Similar with the positive $t_{pp}$ case \cite{Jiang2023PDW}, the spatial oscillation of $n_\alpha(x)$ is also characterized by two ordering wavevectors at $Q$ and $2Q$, corresponding to wavelengths $\lambda_Q\approx 1/\delta$ and $\lambda_{2Q}\approx 1/2\delta$, respectively.

At long distance, the spatial decay of the CDW correlation is dominated by a power-law with an exponent $K_c$, which can be obtained by fitting the charge density oscillations induced by the cylinder boundaries  \cite{White2002}
\begin{eqnarray}\label{Eq:Kc}
n(x)&=& A_Q\ast {\rm cos}(Qx + \phi_1)\ast x^{-K_c/2} \\ 
&+& A_{2Q}\ast {\rm cos}(2Qx + \phi_2)\ast x^{-K_c/2} + n_0.\nonumber
\end{eqnarray}
Here $A_Q$ and $A_{2Q}$ are amplitudes, $\phi_1$ and $\phi_2$ are phase shifts and $n_0$ is the mean density. Examples of the extracted exponents for $\delta=1/8$ are $K_c$(Cu)$\approx 1.6$ and $K_c$(O$_y$)$\approx 1.7$ for $V_{pd}=1$ and $V_{pp}=0.75$, and $K_c$(Cu)$\approx 1.7$ and $K_c$(O$_y$)=1.7 for $V_{pd}=V_{pp}=0$.

\textit{Spin-spin correlations -- }%
To elucidate the magnetic properties of the ground state, we examine the spin-spin correlation functions $F_\alpha(r)=\langle \vec{S}_{x_0,y_0}\cdot \vec{S}_{x_0+r,y_0}\rangle$ where $\alpha$=Cu/O$_x$/O$_y$ site. Fig. \ref{Fig:NxSS}C illustrates examples of $F(r)$, based on two representative parameter choices at $\delta=1/8$. Unlike the dominant spin-spin correlations between Cu sites \cite{Jiang2023PDW}, our findings highlight a dominant correlation between oxygen sites in the kinetically frustrated case. Notably, this decays as a power-law, described by $F(r)\sim r^{-K_s}$ over extended distances. The associated Luttinger exponent is $K_s$(O$_y$)$\approx 1.1$ for $V_{pd}=V_{pp}=0$ and $K_s$(O$_y)\approx 1.3$ for $V_{pd}=-0.8$ and $V_{pp}=-0.2$. In line with the characteristics of a PDW state, $F(r)$ exhibits pronounced spatial oscillations (as seen in the inset of Fig. \ref{Fig:NxSS}C) with a characteristic wavelength, $\lambda_s=1/\delta$. This aligns closely with $\lambda_{sc}$, yielding an ordering wavevector $Q\approx 2\pi\delta$ akin to the SC correlation.

\textit{Entanglement entropy -- }%
Our findings indicate the presence of multiple gapless modes, encompassing both charge and spin degrees of freedom. These can be characterized by the central charge, $c$. This charge is derivable from the von Neumann entropy, formulated as: $S(x)=-{\rm Tr} \rho_x {\rm ln} \rho_x$ where where $\rho_x$ represents the reduced density matrix for a subsystem of length $x$. For critical systems in 1+1 dimensions, described through a conformal field theory, it has been established \cite{Calabrese2004,Fagotti2011} that for an open system of length $L_x$,%
\begin{eqnarray}
S(x)&=&\frac{c}{6} \ln \big[\frac{4(L_x+1)}{\pi} \sin \frac{\pi(2x+1)}{2(L_x+1)}|\sin k_F|\big] \nonumber \\
&+&\tilde{A} \frac{\sin[k_F(2x+1)]}{\frac{4(L_x+1)}{\pi} \sin \frac{\pi(2x+1)}{2(L_x+1)}|\sin k_F|}+ \tilde{S},\label{Eq:EE}
\end{eqnarray}
where $\tilde{A}$ and $\tilde{S}$ are model dependent fitting parameters, and $k_F$ is the Fermi momentum. Our findings reveal a central charge approximately given by $c\approx 2$, with illustrative examples provided in Fig. \ref{Fig:NxSS}D. Specifically, we observed $c\approx 1.95$ at $V_{pd}=V_{pp}=0$ and $c\approx 2.0$ at $V_{pd}=-0.8$ and $V_{pp}=-0.2$ at $\delta=1/8$. These results point to the presence of both a gapless charge mode and a gapless spin mode.

\begin{figure}
  \includegraphics[width=1.0\linewidth]{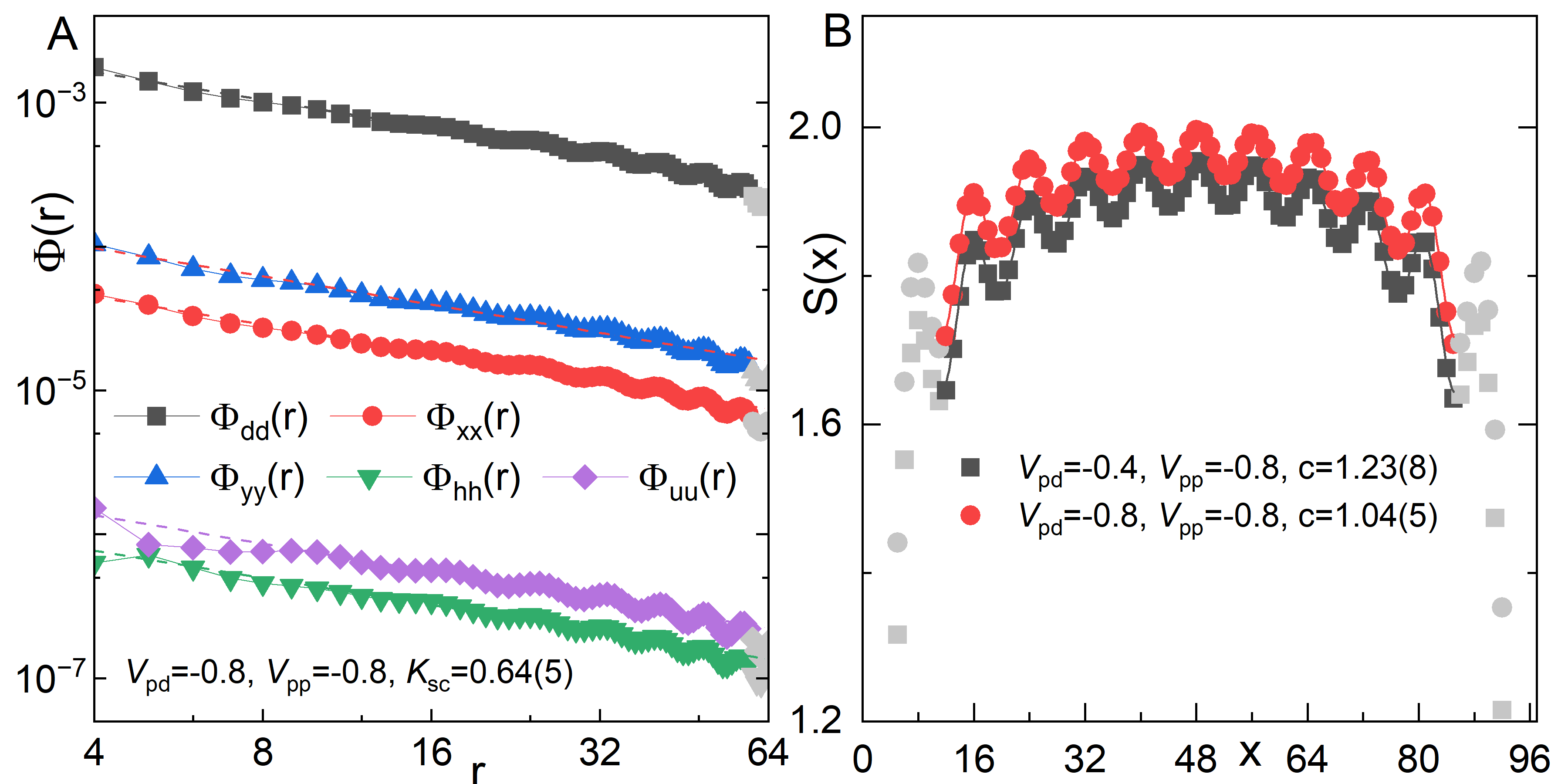}
  \caption{Superconducting correlations and entanglement entropy in the d-wave SC phase at $\delta=1/8$. (A) SC correlations for $V_{pd}=-0.8$ and $V_{pp}=-0.8$. Dashed lines represent fits to a power-law function $f(r)\sim r^{-K_{sc}}$. (B) Von Neumann entanglement entropy $S(x)$. Note that a few data points in gray color close to open ends are excluded to minimize the boundary effect.}\label{Fig:dSC}
\end{figure}

\textbf{$d$-wave superconductivity -- }%
When further increase the attractive $V_{pp}$, the system envolves into a $d$-wave SC phase. Similar with the PDW phase, we find that $\Phi_{dd}(r)$ and $\Phi_{\bar{d}\bar{d}}(r)$ exhibit the strongest SC correlations as shown in Fig. \ref{Fig:dSC}A, i.e. the pairing is dominant between neighboring oxygen sites instead of Cu sites. The pairing symmetry is also consistent with the $d_{xy}$-wave symmetry characterized by the fact $\Phi_{dd}(r)\sim \Phi_{\bar{d}\bar{d}}(r)\sim -\Phi_{d\bar{d}}(r)$.

While there are similarities, several significant distinctions can be drawn between the PDW phase and the $d$-wave SC phase: (1) in the d-wave SC phase, the SC correlations $\Phi_{\alpha\beta}(r)$ maintain a consistent sign in real space, as depicted in Fig. \ref{Fig:dSC}A, and the SC order parameter does not possess finite momentum, (2) the spin-spin correlation functions in this phase are short-ranged and undergo exponential decay, (3) a singular gapless mode with $c\approx 1$ is evident, as illustrated in Fig. \ref{Fig:dSC}B. For instance, the extracted central charge $c\approx 1.04$ for $V_{pd}=-0.8$ and $V_{pp}=-0.8$, and $c\approx 1.2$ for $V_{pd}=-0.8$ and $V_{pp}=-0.4$. Given these observations, our results affirm that the ground state of the $d$-wave SC phase aligns with the characteristics of a Luther-Emery liquid \cite{Emery1987}. This is reminiscent of the single-band Hubbard model on four-leg square cylinders as discussed in prior studies \cite{Jiang2018tJ,Jiang2019Hub,Jiang2020prr,Chung2020,Jiang2020prb,Gong2021,Peng2023}.

\textbf{Summary and discussion -- }%
In conclusion, we have extensively investigated the ground state properties of the lightly doped three-band Hubbard model on two-leg square cylinders, specifically focusing on near-neighbor Cu-O and O-O interactions. Our results strongly suggest that the system's ground state is aligned with the characteristics of a PDW state, showcasing quasi-long-range PDW order and pronounced susceptibility. Several aspects of our findings are unexpected. Within the doped negative $t_{pp}$ Emery model, Cooper pairing prominently emerges between adjacent oxygen sites rather than between neighboring Cu sites. This stands in stark contrast to the prevailing understanding, where Cooper pairing is believed to be dominant between neighboring Cu sites. It has been postulated that cuprate physics can be encapsulated by a single-band effective Hamiltonian, exclusively encompassing the Cu d holes \cite{Zhang1988}. While the pairing symmetry aligns with the $
d$-wave symmetry, it manifests as $d_{xy}$ rather than $d_{x^2-y^2}$. This diverges from the $d_{x^2-y^2}$ pairing symmetry characteristic of cuprates \cite{Lee2006,Fradkin2015}.

\begin{table}[h!]
\centering
\begin{tabular}{| c | c | c |} 
\hline
  & $J$ & $t$ \\ [0.5ex] 
\hline $t_{pp}=-0.5$ & 0.0179 & $-8\times 10^{-4}$ \\ 
\hline
$t_{pp}=0.5$ & 0.165 & -0.673 \\ [1ex] 
\hline
\end{tabular}
\caption{Effective single band exchange parameter $J$ and NN hopping $t$ determined from Cu$_2$O$_7$ clusters for different values of $t_{pp}$ as indicated. All other parameters are the same as used in the main text.}
\label{table:1}
\end{table}

To understand these results, we return to cluster calculations that determine the relevant parameters $t$ and $J$ of an effective single band model \cite{Eskes1990,Eskes1989}. Specifically we consider Cu$_2$O$_7$ clusters with the same parameters used for DMRG, and compare the results for positive and negative $t_{pp}$. Our results are summarized in Table \ref{table:1}. Determining the singlet-triplet energy difference for two holes on the cluster yields an exchange energy $J=0.0179$ for $t_{pp}=-0.5$, compared to 0.165 if the sign of $t_{pp}$ is reversed. Largely interpreted as due to an increase in the effective charge transfer energy when ligand delocalization is considered, the effective spin exchange between Cu spins is greatly reduced for negative $t_{pp}$. While the dependence on $V_{pp}$ is negligible, a negative $V_{pd}$ increases the magnetic exchange for the two-hole ground state configuration. The increase of $J$ for negative $V_{pd}$ may help to favor stronger hole singlet bonding, promoting stronger SC susceptibilities in Fig. \ref{Fig:PDW}.

\begin{figure}[!th]
\centering
  \includegraphics[width=1\linewidth]{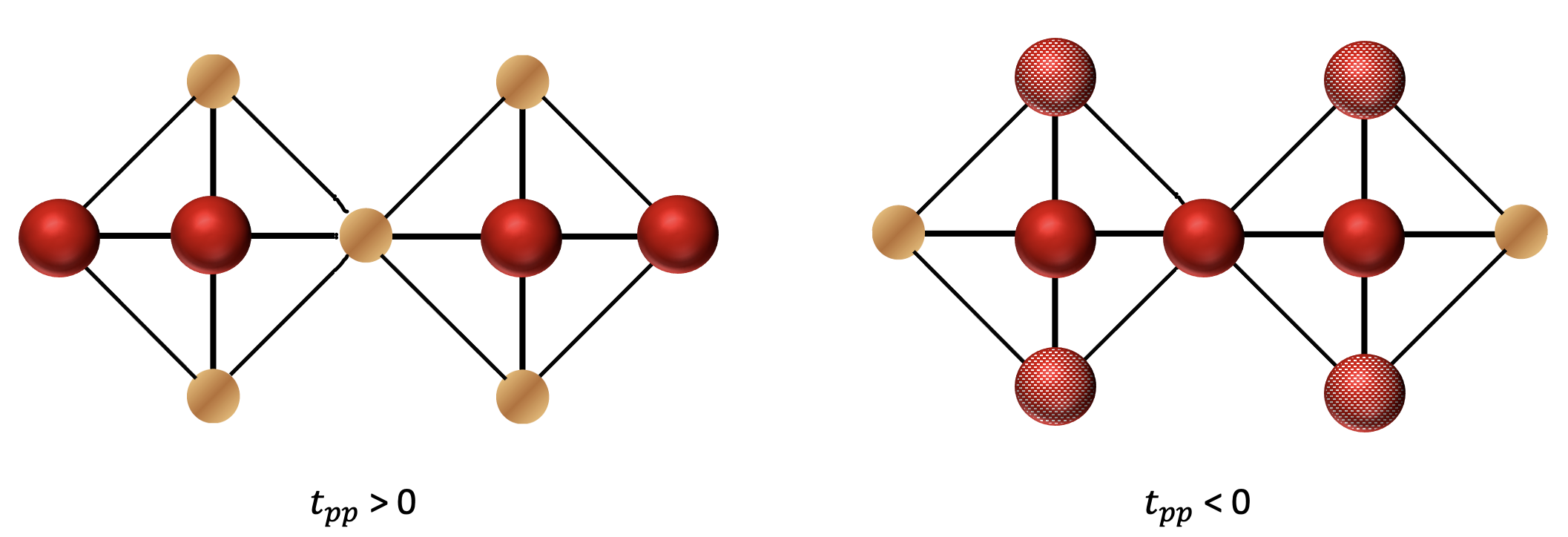}
    \caption{Predominant hole distributions for $t_{pp}$ positive or negative, as indicated. Solid red spheres denote approximately a full hole charge, while shaded red spheres denote an approximate quarter charge.}\label{Fig:holes}
\end{figure}

Calculations for three holes on the same cluster yields the hopping parameter $t$, defined as the energy difference between the ground and first excited state. For $V_{pp}=V_{pd}=0$, $2t=-0.673$ for $t_{pp}=0.5$ while $2t=-8\times 10^{-4}$ for $t_{pp}=-0.5$. These numbers increase slightly for $V_{pp}<0$, but overall a reversed sign of $t_{pp}$ dramatically affects the magnitude of the NN hopping.

Lastly, binding of doped holes can be examined for the ground state of four doped holes on the same cluster (see Fig. \ref{Fig:holes}). For both $t_{pp}$ positive and negative, the ground state is a spin-singlet, with Cu spins and O spins both forming singlets. However, the spatial orientation of bound holes on O is different: for $t_{pp}=0.5$, O holes primarily bind to Cu at the ends of the cluster, without substantial hole occupation on the central, bridging oxygen, while for $t_{pp}=-0.5$, O holes primarily bind to the central oxygen in a local $d_{xy}$ configuration with neighboring oxygens. As this might be expected when antiferromagnetic exchange among oxygen becomes dominant over Cu, the predominance of $d_{xy}$ pairing observed in the phase diagram of Fig. \ref{Fig:Phase} may be related to this tendency to bind neighboring O holes.



\textbf{Acknowledgements -- }%
We are grateful to Steven Kivelson for insightful discussions and invaluable suggestions. This work was supported by the Department of Energy, Office of Science, Basic Energy Sciences, Materials Sciences and Engineering Division, under Contract DE-AC02-76SF00515.



\begin{thebibliography}{60}
\expandafter\ifx\csname natexlab\endcsname\relax\def\natexlab#1{#1}\fi
\expandafter\ifx\csname bibnamefont\endcsname\relax
  \def\bibnamefont#1{#1}\fi
\expandafter\ifx\csname bibfnamefont\endcsname\relax
  \def\bibfnamefont#1{#1}\fi
\expandafter\ifx\csname citenamefont\endcsname\relax
  \def\citenamefont#1{#1}\fi
\expandafter\ifx\csname url\endcsname\relax
  \def\url#1{\texttt{#1}}\fi
\expandafter\ifx\csname urlprefix\endcsname\relax\def\urlprefix{URL }\fi
\providecommand{\bibinfo}[2]{#2}
\providecommand{\eprint}[2][]{\url{#2}}

\bibitem[{\citenamefont{Zaanen et~al.}(1985)\citenamefont{Zaanen, Sawatzky, and
  Allen}}]{Zaanen1985}
\bibinfo{author}{\bibfnamefont{J.}~\bibnamefont{Zaanen}},
  \bibinfo{author}{\bibfnamefont{G.~A.} \bibnamefont{Sawatzky}},
  \bibnamefont{and} \bibinfo{author}{\bibfnamefont{J.~W.} \bibnamefont{Allen}},
  \bibinfo{journal}{Phys. Rev. Lett.} \textbf{\bibinfo{volume}{55}},
  \bibinfo{pages}{418} (\bibinfo{year}{1985}),
  \urlprefix\url{https://link.aps.org/doi/10.1103/PhysRevLett.55.418}.

\bibitem[{\citenamefont{Emery}(1987)}]{Emery1987}
\bibinfo{author}{\bibfnamefont{V.~J.} \bibnamefont{Emery}},
  \bibinfo{journal}{Phys. Rev. Lett.} \textbf{\bibinfo{volume}{58}},
  \bibinfo{pages}{2794} (\bibinfo{year}{1987}),
  \urlprefix\url{https://link.aps.org/doi/10.1103/PhysRevLett.58.2794}.

\bibitem[{\citenamefont{Scalettar}(1989)}]{Scalettar1989}
\bibinfo{author}{\bibfnamefont{R.}~\bibnamefont{Scalettar}},
  \bibinfo{journal}{Physica C: Superconductivity and its Applications}
  \textbf{\bibinfo{volume}{162-164}}, \bibinfo{pages}{313}
  (\bibinfo{year}{1989}), ISSN \bibinfo{issn}{0921-4534},
  \urlprefix\url{https://www.sciencedirect.com/science/article/pii/0921453489910332}.

\bibitem[{\citenamefont{Scalettar et~al.}(1991)\citenamefont{Scalettar,
  Scalapino, Sugar, and White}}]{Scalettar1991}
\bibinfo{author}{\bibfnamefont{R.~T.} \bibnamefont{Scalettar}},
  \bibinfo{author}{\bibfnamefont{D.~J.} \bibnamefont{Scalapino}},
  \bibinfo{author}{\bibfnamefont{R.~L.} \bibnamefont{Sugar}}, \bibnamefont{and}
  \bibinfo{author}{\bibfnamefont{S.~R.} \bibnamefont{White}},
  \bibinfo{journal}{Phys. Rev. B} \textbf{\bibinfo{volume}{44}},
  \bibinfo{pages}{770} (\bibinfo{year}{1991}),
  \urlprefix\url{https://link.aps.org/doi/10.1103/PhysRevB.44.770}.

\bibitem[{\citenamefont{White and Scalapino}(2015)}]{White2015}
\bibinfo{author}{\bibfnamefont{S.~R.} \bibnamefont{White}} \bibnamefont{and}
  \bibinfo{author}{\bibfnamefont{D.~J.} \bibnamefont{Scalapino}},
  \bibinfo{journal}{Phys. Rev. B} \textbf{\bibinfo{volume}{92}},
  \bibinfo{pages}{205112} (\bibinfo{year}{2015}),
  \urlprefix\url{https://link.aps.org/doi/10.1103/PhysRevB.92.205112}.

\bibitem[{\citenamefont{Huang et~al.}(2017)\citenamefont{Huang, Mendl, Liu,
  Johnston, Jiang, Moritz, and Devereaux}}]{Huang2017}
\bibinfo{author}{\bibfnamefont{E.~W.} \bibnamefont{Huang}},
  \bibinfo{author}{\bibfnamefont{C.~B.} \bibnamefont{Mendl}},
  \bibinfo{author}{\bibfnamefont{S.}~\bibnamefont{Liu}},
  \bibinfo{author}{\bibfnamefont{S.}~\bibnamefont{Johnston}},
  \bibinfo{author}{\bibfnamefont{H.-C.} \bibnamefont{Jiang}},
  \bibinfo{author}{\bibfnamefont{B.}~\bibnamefont{Moritz}}, \bibnamefont{and}
  \bibinfo{author}{\bibfnamefont{T.~P.} \bibnamefont{Devereaux}},
  \bibinfo{journal}{Science} \textbf{\bibinfo{volume}{358}},
  \bibinfo{pages}{1161} (\bibinfo{year}{2017}),
  \eprint{https://www.science.org/doi/pdf/10.1126/science.aak9546},
  \urlprefix\url{https://www.science.org/doi/abs/10.1126/science.aak9546}.

\bibitem[{\citenamefont{Jiang et~al.}(2023)\citenamefont{Jiang, Scalapino, and
  White}}]{Jiang2023DF}
\bibinfo{author}{\bibfnamefont{S.}~\bibnamefont{Jiang}},
  \bibinfo{author}{\bibfnamefont{D.~J.} \bibnamefont{Scalapino}},
  \bibnamefont{and} \bibinfo{author}{\bibfnamefont{S.~R.} \bibnamefont{White}}
  (\bibinfo{year}{2023}), \eprint{arXiv:2303.00756}.

\bibitem[{\citenamefont{Jiang}(2023)}]{Jiang2023PDW}
\bibinfo{author}{\bibfnamefont{H.-C.} \bibnamefont{Jiang}},
  \bibinfo{journal}{Phys. Rev. B} \textbf{\bibinfo{volume}{107}},
  \bibinfo{pages}{214504} (\bibinfo{year}{2023}),
  \urlprefix\url{https://link.aps.org/doi/10.1103/PhysRevB.107.214504}.

\bibitem[{\citenamefont{Fulde and Ferrell}(1964)}]{FF1964}
\bibinfo{author}{\bibfnamefont{P.}~\bibnamefont{Fulde}} \bibnamefont{and}
  \bibinfo{author}{\bibfnamefont{R.~A.} \bibnamefont{Ferrell}},
  \bibinfo{journal}{Phys. Rev.} \textbf{\bibinfo{volume}{135}},
  \bibinfo{pages}{A550} (\bibinfo{year}{1964}).

\bibitem[{\citenamefont{Larkin and Ovchinnikov}(1965)}]{LO1965}
\bibinfo{author}{\bibfnamefont{A.~I.} \bibnamefont{Larkin}} \bibnamefont{and}
  \bibinfo{author}{\bibfnamefont{Y.~N.} \bibnamefont{Ovchinnikov}},
  \bibinfo{journal}{Sov. Phys. JETP} \textbf{\bibinfo{volume}{20}},
  \bibinfo{pages}{762} (\bibinfo{year}{1965}).

\bibitem[{\citenamefont{Berg et~al.}(2009)\citenamefont{Berg, Fradkin,
  Kivelson, and Tranquada}}]{Berg2009}
\bibinfo{author}{\bibfnamefont{E.}~\bibnamefont{Berg}},
  \bibinfo{author}{\bibfnamefont{E.}~\bibnamefont{Fradkin}},
  \bibinfo{author}{\bibfnamefont{S.~A.} \bibnamefont{Kivelson}},
  \bibnamefont{and} \bibinfo{author}{\bibfnamefont{J.~M.}
  \bibnamefont{Tranquada}}, \bibinfo{journal}{New J. Phys.}
  \textbf{\bibinfo{volume}{11}}, \bibinfo{pages}{115004}
  (\bibinfo{year}{2009}).

\bibitem[{\citenamefont{Fradkin et~al.}(2015)\citenamefont{Fradkin, Kivelson,
  and Tranquada}}]{Fradkin2015}
\bibinfo{author}{\bibfnamefont{E.}~\bibnamefont{Fradkin}},
  \bibinfo{author}{\bibfnamefont{S.~A.} \bibnamefont{Kivelson}},
  \bibnamefont{and} \bibinfo{author}{\bibfnamefont{J.~M.}
  \bibnamefont{Tranquada}}, \bibinfo{journal}{Rev. Mod. Phys.}
  \textbf{\bibinfo{volume}{87}}, \bibinfo{pages}{457} (\bibinfo{year}{2015}),
  \urlprefix\url{https://link.aps.org/doi/10.1103/RevModPhys.87.457}.

\bibitem[{\citenamefont{Agterberg et~al.}(2020)\citenamefont{Agterberg, Davis,
  Edkins, Fradkin, Van~Harlingen, Kivelson, Lee, Radzihovsky, Tranquada, and
  Wang}}]{Agterberg2020}
\bibinfo{author}{\bibfnamefont{D.~F.} \bibnamefont{Agterberg}},
  \bibinfo{author}{\bibfnamefont{J.~S.} \bibnamefont{Davis}},
  \bibinfo{author}{\bibfnamefont{S.~D.} \bibnamefont{Edkins}},
  \bibinfo{author}{\bibfnamefont{E.}~\bibnamefont{Fradkin}},
  \bibinfo{author}{\bibfnamefont{D.~J.} \bibnamefont{Van~Harlingen}},
  \bibinfo{author}{\bibfnamefont{S.~A.} \bibnamefont{Kivelson}},
  \bibinfo{author}{\bibfnamefont{P.~A.} \bibnamefont{Lee}},
  \bibinfo{author}{\bibfnamefont{L.}~\bibnamefont{Radzihovsky}},
  \bibinfo{author}{\bibfnamefont{J.~M.} \bibnamefont{Tranquada}},
  \bibnamefont{and} \bibinfo{author}{\bibfnamefont{Y.}~\bibnamefont{Wang}},
  \bibinfo{journal}{Annual Review of Condensed Matter Physics}
  \textbf{\bibinfo{volume}{11}}, \bibinfo{pages}{231} (\bibinfo{year}{2020}),
  \bibinfo{note}{and references therein.}

\bibitem[{\citenamefont{Lee}(2014)}]{Lee2014}
\bibinfo{author}{\bibfnamefont{P.~A.} \bibnamefont{Lee}},
  \bibinfo{journal}{Phys. Rev. X} \textbf{\bibinfo{volume}{4}},
  \bibinfo{pages}{031017} (\bibinfo{year}{2014}).

\bibitem[{\citenamefont{Jian et~al.}(2020)\citenamefont{Jian, Scherer, and
  Yao}}]{Jian2020}
\bibinfo{author}{\bibfnamefont{S.-K.} \bibnamefont{Jian}},
  \bibinfo{author}{\bibfnamefont{M.~M.} \bibnamefont{Scherer}},
  \bibnamefont{and} \bibinfo{author}{\bibfnamefont{H.}~\bibnamefont{Yao}},
  \bibinfo{journal}{Phys. Rev. Research} \textbf{\bibinfo{volume}{2}},
  \bibinfo{pages}{013034} (\bibinfo{year}{2020}).

\bibitem[{\citenamefont{Lozano et~al.}(2022)\citenamefont{Lozano, Ren, Gu,
  Tsvelik, Tranquada, and Li}}]{Lozano2022}
\bibinfo{author}{\bibfnamefont{P.~M.} \bibnamefont{Lozano}},
  \bibinfo{author}{\bibfnamefont{T.}~\bibnamefont{Ren}},
  \bibinfo{author}{\bibfnamefont{G.~D.} \bibnamefont{Gu}},
  \bibinfo{author}{\bibfnamefont{A.~M.} \bibnamefont{Tsvelik}},
  \bibinfo{author}{\bibfnamefont{J.~M.} \bibnamefont{Tranquada}},
  \bibnamefont{and} \bibinfo{author}{\bibfnamefont{Q.}~\bibnamefont{Li}},
  \bibinfo{journal}{Phys. Rev. B} \textbf{\bibinfo{volume}{106}},
  \bibinfo{pages}{174510} (\bibinfo{year}{2022}),
  \urlprefix\url{https://link.aps.org/doi/10.1103/PhysRevB.106.174510}.

\bibitem[{\citenamefont{Himeda et~al.}(2002)\citenamefont{Himeda, Kato, and
  Ogata}}]{Himeda2002}
\bibinfo{author}{\bibfnamefont{A.}~\bibnamefont{Himeda}},
  \bibinfo{author}{\bibfnamefont{T.}~\bibnamefont{Kato}}, \bibnamefont{and}
  \bibinfo{author}{\bibfnamefont{M.}~\bibnamefont{Ogata}},
  \bibinfo{journal}{Phys. Rev. Lett.} \textbf{\bibinfo{volume}{88}},
  \bibinfo{pages}{117001} (\bibinfo{year}{2002}).

\bibitem[{\citenamefont{Hamidian et~al.}(2016)\citenamefont{Hamidian, Edkins,
  Joo, Kostin, Eisaki, Uchida, Lawler, Kim, Mackenzie, Fujita
  et~al.}}]{Hamidian2016}
\bibinfo{author}{\bibfnamefont{M.~H.} \bibnamefont{Hamidian}},
  \bibinfo{author}{\bibfnamefont{S.~D.} \bibnamefont{Edkins}},
  \bibinfo{author}{\bibfnamefont{S.~H.} \bibnamefont{Joo}},
  \bibinfo{author}{\bibfnamefont{A.}~\bibnamefont{Kostin}},
  \bibinfo{author}{\bibfnamefont{H.}~\bibnamefont{Eisaki}},
  \bibinfo{author}{\bibfnamefont{S.}~\bibnamefont{Uchida}},
  \bibinfo{author}{\bibfnamefont{M.~J.} \bibnamefont{Lawler}},
  \bibinfo{author}{\bibfnamefont{E.-A.} \bibnamefont{Kim}},
  \bibinfo{author}{\bibfnamefont{A.~P.} \bibnamefont{Mackenzie}},
  \bibinfo{author}{\bibfnamefont{K.}~\bibnamefont{Fujita}},
  \bibnamefont{et~al.}, \bibinfo{journal}{Nature}
  \textbf{\bibinfo{volume}{532}}, \bibinfo{pages}{343} (\bibinfo{year}{2016}).

\bibitem[{\citenamefont{Ruan et~al.}(2018)\citenamefont{Ruan, Li, Hu, Hao, Li,
  Cai, Zhou, Lee, and Wang}}]{Ruan2018}
\bibinfo{author}{\bibfnamefont{W.}~\bibnamefont{Ruan}},
  \bibinfo{author}{\bibfnamefont{X.}~\bibnamefont{Li}},
  \bibinfo{author}{\bibfnamefont{C.}~\bibnamefont{Hu}},
  \bibinfo{author}{\bibfnamefont{Z.}~\bibnamefont{Hao}},
  \bibinfo{author}{\bibfnamefont{H.}~\bibnamefont{Li}},
  \bibinfo{author}{\bibfnamefont{P.}~\bibnamefont{Cai}},
  \bibinfo{author}{\bibfnamefont{X.}~\bibnamefont{Zhou}},
  \bibinfo{author}{\bibfnamefont{D.-H.} \bibnamefont{Lee}}, \bibnamefont{and}
  \bibinfo{author}{\bibfnamefont{Y.}~\bibnamefont{Wang}},
  \bibinfo{journal}{Nature Physics} pp. \bibinfo{pages}{1178--1182}
  (\bibinfo{year}{2018}), ISSN \bibinfo{issn}{1745-2481}.

\bibitem[{\citenamefont{Edkins et~al.}(2019)\citenamefont{Edkins, Kostin,
  Fujita, Mackenzie, Eisaki, Uchida, Sachdev, Lawler, Kim, S{\'e}amus~Davis
  et~al.}}]{Edkins2019}
\bibinfo{author}{\bibfnamefont{S.~D.} \bibnamefont{Edkins}},
  \bibinfo{author}{\bibfnamefont{A.}~\bibnamefont{Kostin}},
  \bibinfo{author}{\bibfnamefont{K.}~\bibnamefont{Fujita}},
  \bibinfo{author}{\bibfnamefont{A.~P.} \bibnamefont{Mackenzie}},
  \bibinfo{author}{\bibfnamefont{H.}~\bibnamefont{Eisaki}},
  \bibinfo{author}{\bibfnamefont{S.}~\bibnamefont{Uchida}},
  \bibinfo{author}{\bibfnamefont{S.}~\bibnamefont{Sachdev}},
  \bibinfo{author}{\bibfnamefont{M.~J.} \bibnamefont{Lawler}},
  \bibinfo{author}{\bibfnamefont{E.-A.} \bibnamefont{Kim}},
  \bibinfo{author}{\bibfnamefont{J.~C.} \bibnamefont{S{\'e}amus~Davis}},
  \bibnamefont{et~al.}, \bibinfo{journal}{Science}
  \textbf{\bibinfo{volume}{364}}, \bibinfo{pages}{976} (\bibinfo{year}{2019}),
  ISSN \bibinfo{issn}{0036-8075}.

\bibitem[{\citenamefont{Liu et~al.}(2021)\citenamefont{Liu, Chong, Sharma, and
  Davis}}]{Liu2021}
\bibinfo{author}{\bibfnamefont{X.}~\bibnamefont{Liu}},
  \bibinfo{author}{\bibfnamefont{Y.~X.} \bibnamefont{Chong}},
  \bibinfo{author}{\bibfnamefont{R.}~\bibnamefont{Sharma}}, \bibnamefont{and}
  \bibinfo{author}{\bibfnamefont{J.~C.~S.} \bibnamefont{Davis}},
  \bibinfo{journal}{Science} \textbf{\bibinfo{volume}{372}},
  \bibinfo{pages}{1447} (\bibinfo{year}{2021}),
  \eprint{https://www.science.org/doi/pdf/10.1126/science.abd4607},
  \urlprefix\url{https://www.science.org/doi/abs/10.1126/science.abd4607}.

\bibitem[{\citenamefont{Agterberg and Tsunetsugu}(2008)}]{Agterberg2008}
\bibinfo{author}{\bibfnamefont{D.~F.} \bibnamefont{Agterberg}}
  \bibnamefont{and}
  \bibinfo{author}{\bibfnamefont{H.}~\bibnamefont{Tsunetsugu}},
  \bibinfo{journal}{Nature Physics} pp. \bibinfo{pages}{639--642}
  (\bibinfo{year}{2008}), ISSN \bibinfo{issn}{1745-2481}.

\bibitem[{\citenamefont{Li et~al.}(2007)\citenamefont{Li, H\"ucker, Gu,
  Tsvelik, and Tranquada}}]{Li2007}
\bibinfo{author}{\bibfnamefont{Q.}~\bibnamefont{Li}},
  \bibinfo{author}{\bibfnamefont{M.}~\bibnamefont{H\"ucker}},
  \bibinfo{author}{\bibfnamefont{G.~D.} \bibnamefont{Gu}},
  \bibinfo{author}{\bibfnamefont{A.~M.} \bibnamefont{Tsvelik}},
  \bibnamefont{and} \bibinfo{author}{\bibfnamefont{J.~M.}
  \bibnamefont{Tranquada}}, \bibinfo{journal}{Phys. Rev. Lett.}
  \textbf{\bibinfo{volume}{99}}, \bibinfo{pages}{067001}
  (\bibinfo{year}{2007}),
  \urlprefix\url{https://link.aps.org/doi/10.1103/PhysRevLett.99.067001}.

\bibitem[{\citenamefont{Berg et~al.}(2007)\citenamefont{Berg, Fradkin, Kim,
  Kivelson, Oganesyan, Tranquada, and Zhang}}]{Berg2007}
\bibinfo{author}{\bibfnamefont{E.}~\bibnamefont{Berg}},
  \bibinfo{author}{\bibfnamefont{E.}~\bibnamefont{Fradkin}},
  \bibinfo{author}{\bibfnamefont{E.-A.} \bibnamefont{Kim}},
  \bibinfo{author}{\bibfnamefont{S.~A.} \bibnamefont{Kivelson}},
  \bibinfo{author}{\bibfnamefont{V.}~\bibnamefont{Oganesyan}},
  \bibinfo{author}{\bibfnamefont{J.~M.} \bibnamefont{Tranquada}},
  \bibnamefont{and} \bibinfo{author}{\bibfnamefont{S.~C.} \bibnamefont{Zhang}},
  \bibinfo{journal}{Phys. Rev. Lett.} \textbf{\bibinfo{volume}{99}},
  \bibinfo{pages}{127003} (\bibinfo{year}{2007}).

\bibitem[{\citenamefont{Tranquada et~al.}(2008)\citenamefont{Tranquada, Gu,
  H\"ucker, Jie, Kang, Klingeler, Li, Tristan, Wen, Xu et~al.}}]{Tranquada2008}
\bibinfo{author}{\bibfnamefont{J.~M.} \bibnamefont{Tranquada}},
  \bibinfo{author}{\bibfnamefont{G.~D.} \bibnamefont{Gu}},
  \bibinfo{author}{\bibfnamefont{M.}~\bibnamefont{H\"ucker}},
  \bibinfo{author}{\bibfnamefont{Q.}~\bibnamefont{Jie}},
  \bibinfo{author}{\bibfnamefont{H.-J.} \bibnamefont{Kang}},
  \bibinfo{author}{\bibfnamefont{R.}~\bibnamefont{Klingeler}},
  \bibinfo{author}{\bibfnamefont{Q.}~\bibnamefont{Li}},
  \bibinfo{author}{\bibfnamefont{N.}~\bibnamefont{Tristan}},
  \bibinfo{author}{\bibfnamefont{J.~S.} \bibnamefont{Wen}},
  \bibinfo{author}{\bibfnamefont{G.~Y.} \bibnamefont{Xu}},
  \bibnamefont{et~al.}, \bibinfo{journal}{Phys. Rev. B}
  \textbf{\bibinfo{volume}{78}}, \bibinfo{pages}{174529}
  (\bibinfo{year}{2008}),
  \urlprefix\url{https://link.aps.org/doi/10.1103/PhysRevB.78.174529}.

\bibitem[{\citenamefont{Tranquada}(2020)}]{Tranquada2020}
\bibinfo{author}{\bibfnamefont{J.~M.} \bibnamefont{Tranquada}},
  \bibinfo{journal}{Advances in Physics} \textbf{\bibinfo{volume}{69}},
  \bibinfo{pages}{437} (\bibinfo{year}{2020}),
  \eprint{https://doi.org/10.1080/00018732.2021.1935698},
  \urlprefix\url{https://doi.org/10.1080/00018732.2021.1935698}.

\bibitem[{\citenamefont{Tranquada}(2021)}]{Tranquada2021}
\bibinfo{author}{\bibfnamefont{J.~M.} \bibnamefont{Tranquada}},
  \bibinfo{journal}{Symmetry} \textbf{\bibinfo{volume}{13}}
  (\bibinfo{year}{2021}), ISSN \bibinfo{issn}{2073-8994},
  \urlprefix\url{https://www.mdpi.com/2073-8994/13/12/2365}.

\bibitem[{\citenamefont{Chen et~al.}(2021{\natexlab{a}})\citenamefont{Chen,
  Yang, Hu, Zhao, Yuan, Xing, Qian, Huang, Li, Ye et~al.}}]{Chen2021k}
\bibinfo{author}{\bibfnamefont{H.}~\bibnamefont{Chen}},
  \bibinfo{author}{\bibfnamefont{H.}~\bibnamefont{Yang}},
  \bibinfo{author}{\bibfnamefont{B.}~\bibnamefont{Hu}},
  \bibinfo{author}{\bibfnamefont{Z.}~\bibnamefont{Zhao}},
  \bibinfo{author}{\bibfnamefont{J.}~\bibnamefont{Yuan}},
  \bibinfo{author}{\bibfnamefont{Y.}~\bibnamefont{Xing}},
  \bibinfo{author}{\bibfnamefont{G.}~\bibnamefont{Qian}},
  \bibinfo{author}{\bibfnamefont{Z.}~\bibnamefont{Huang}},
  \bibinfo{author}{\bibfnamefont{G.}~\bibnamefont{Li}},
  \bibinfo{author}{\bibfnamefont{Y.}~\bibnamefont{Ye}}, \bibnamefont{et~al.},
  \bibinfo{journal}{Nature} \textbf{\bibinfo{volume}{599}},
  \bibinfo{pages}{222–228} (\bibinfo{year}{2021}{\natexlab{a}}).

\bibitem[{\citenamefont{Liu et~al.}(2023)\citenamefont{Liu, Wei, He, Zhang,
  Wang, and Wang}}]{Liu2023}
\bibinfo{author}{\bibfnamefont{Y.}~\bibnamefont{Liu}},
  \bibinfo{author}{\bibfnamefont{T.}~\bibnamefont{Wei}},
  \bibinfo{author}{\bibfnamefont{G.}~\bibnamefont{He}},
  \bibinfo{author}{\bibfnamefont{Y.}~\bibnamefont{Zhang}},
  \bibinfo{author}{\bibfnamefont{Z.}~\bibnamefont{Wang}}, \bibnamefont{and}
  \bibinfo{author}{\bibfnamefont{J.}~\bibnamefont{Wang}},
  \bibinfo{journal}{Nature} \textbf{\bibinfo{volume}{618}},
  \bibinfo{pages}{934–939} (\bibinfo{year}{2023}).

\bibitem[{\citenamefont{Berg et~al.}(2010)\citenamefont{Berg, Fradkin, and
  Kivelson}}]{Berg2010}
\bibinfo{author}{\bibfnamefont{E.}~\bibnamefont{Berg}},
  \bibinfo{author}{\bibfnamefont{E.}~\bibnamefont{Fradkin}}, \bibnamefont{and}
  \bibinfo{author}{\bibfnamefont{S.~A.} \bibnamefont{Kivelson}},
  \bibinfo{journal}{Phys. Rev. Lett.} \textbf{\bibinfo{volume}{105}},
  \bibinfo{pages}{146403} (\bibinfo{year}{2010}).

\bibitem[{\citenamefont{Jaefari and Fradkin}(2012)}]{Fradkin2012}
\bibinfo{author}{\bibfnamefont{A.}~\bibnamefont{Jaefari}} \bibnamefont{and}
  \bibinfo{author}{\bibfnamefont{E.}~\bibnamefont{Fradkin}},
  \bibinfo{journal}{Phys. Rev. B} \textbf{\bibinfo{volume}{85}},
  \bibinfo{pages}{035104} (\bibinfo{year}{2012}).

\bibitem[{\citenamefont{Venderley and Kim}(2019)}]{Venderley2019}
\bibinfo{author}{\bibfnamefont{J.}~\bibnamefont{Venderley}} \bibnamefont{and}
  \bibinfo{author}{\bibfnamefont{E.-A.} \bibnamefont{Kim}},
  \bibinfo{journal}{Science Advances} \textbf{\bibinfo{volume}{5}},
  \bibinfo{pages}{eaat4698} (\bibinfo{year}{2019}).

\bibitem[{\citenamefont{Xu et~al.}(2019)\citenamefont{Xu, Law, and
  Lee}}]{Xu2019}
\bibinfo{author}{\bibfnamefont{X.~Y.} \bibnamefont{Xu}},
  \bibinfo{author}{\bibfnamefont{K.~T.} \bibnamefont{Law}}, \bibnamefont{and}
  \bibinfo{author}{\bibfnamefont{P.~A.} \bibnamefont{Lee}},
  \bibinfo{journal}{Phys. Rev. Lett.} \textbf{\bibinfo{volume}{122}},
  \bibinfo{pages}{167001} (\bibinfo{year}{2019}).

\bibitem[{\citenamefont{Peng et~al.}(2021{\natexlab{a}})\citenamefont{Peng,
  Jiang, Devereaux, and Jiang}}]{Peng2021a}
\bibinfo{author}{\bibfnamefont{C.}~\bibnamefont{Peng}},
  \bibinfo{author}{\bibfnamefont{Y.-F.} \bibnamefont{Jiang}},
  \bibinfo{author}{\bibfnamefont{T.~P.} \bibnamefont{Devereaux}},
  \bibnamefont{and} \bibinfo{author}{\bibfnamefont{H.-C.} \bibnamefont{Jiang}},
  \bibinfo{journal}{npj Quantum Mater.} \textbf{\bibinfo{volume}{6}},
  \bibinfo{pages}{64} (\bibinfo{year}{2021}{\natexlab{a}}).

\bibitem[{\citenamefont{Peng et~al.}(2021{\natexlab{b}})\citenamefont{Peng,
  Jiang, Wang, and Jiang}}]{Peng2021b}
\bibinfo{author}{\bibfnamefont{C.}~\bibnamefont{Peng}},
  \bibinfo{author}{\bibfnamefont{Y.-F.} \bibnamefont{Jiang}},
  \bibinfo{author}{\bibfnamefont{Y.}~\bibnamefont{Wang}}, \bibnamefont{and}
  \bibinfo{author}{\bibfnamefont{H.-C.} \bibnamefont{Jiang}},
  \bibinfo{journal}{New Journal of Physics} \textbf{\bibinfo{volume}{23}},
  \bibinfo{pages}{123004} (\bibinfo{year}{2021}{\natexlab{b}}),
  \urlprefix\url{https://doi.org/10.1088/1367-2630/ac3a83}.

\bibitem[{\citenamefont{Han et~al.}(2020)\citenamefont{Han, Kivelson, and
  Yao}}]{Han2020}
\bibinfo{author}{\bibfnamefont{Z.}~\bibnamefont{Han}},
  \bibinfo{author}{\bibfnamefont{S.~A.} \bibnamefont{Kivelson}},
  \bibnamefont{and} \bibinfo{author}{\bibfnamefont{H.}~\bibnamefont{Yao}},
  \bibinfo{journal}{Phys. Rev. Lett.} \textbf{\bibinfo{volume}{125}},
  \bibinfo{pages}{167001} (\bibinfo{year}{2020}).

\bibitem[{\citenamefont{Huang et~al.}(2022)\citenamefont{Huang, Han, Kivelson,
  and Yao}}]{Huang2022}
\bibinfo{author}{\bibfnamefont{K.~S.} \bibnamefont{Huang}},
  \bibinfo{author}{\bibfnamefont{Z.}~\bibnamefont{Han}},
  \bibinfo{author}{\bibfnamefont{S.~A.} \bibnamefont{Kivelson}},
  \bibnamefont{and} \bibinfo{author}{\bibfnamefont{H.}~\bibnamefont{Yao}},
  \bibinfo{journal}{npj Quantum Mater.} \textbf{\bibinfo{volume}{7}},
  \bibinfo{pages}{17} (\bibinfo{year}{2022}).

\bibitem[{\citenamefont{Wu et~al.}(2023)\citenamefont{Wu, Nosov, Patel, and
  Raghu}}]{Wu2023}
\bibinfo{author}{\bibfnamefont{Y.-M.} \bibnamefont{Wu}},
  \bibinfo{author}{\bibfnamefont{P.~A.} \bibnamefont{Nosov}},
  \bibinfo{author}{\bibfnamefont{A.~A.} \bibnamefont{Patel}}, \bibnamefont{and}
  \bibinfo{author}{\bibfnamefont{S.}~\bibnamefont{Raghu}},
  \bibinfo{journal}{Phys. Rev. Lett.} \textbf{\bibinfo{volume}{130}},
  \bibinfo{pages}{026001} (\bibinfo{year}{2023}),
  \urlprefix\url{https://link.aps.org/doi/10.1103/PhysRevLett.130.026001}.

\bibitem[{\citenamefont{Jiang and Yao}(2023)}]{Jiang2023YF}
\bibinfo{author}{\bibfnamefont{Y.-F.} \bibnamefont{Jiang}} \bibnamefont{and}
  \bibinfo{author}{\bibfnamefont{H.}~\bibnamefont{Yao}} (\bibinfo{year}{2023}),
  \eprint{arXiv:2308.08609}.

\bibitem[{\citenamefont{White}(1992)}]{White1992}
\bibinfo{author}{\bibfnamefont{S.~R.} \bibnamefont{White}},
  \bibinfo{journal}{Phys. Rev. Lett.} \textbf{\bibinfo{volume}{69}},
  \bibinfo{pages}{2863} (\bibinfo{year}{1992}).

\bibitem[{\citenamefont{White}(1993)}]{White1993}
\bibinfo{author}{\bibfnamefont{S.~R.} \bibnamefont{White}},
  \bibinfo{journal}{Phys. Rev. B} \textbf{\bibinfo{volume}{48}},
  \bibinfo{pages}{10345} (\bibinfo{year}{1993}),
  \urlprefix\url{https://link.aps.org/doi/10.1103/PhysRevB.48.10345}.

\bibitem[{\citenamefont{Eskes et~al.}(1990)\citenamefont{Eskes, Tjeng, and
  Sawatzky}}]{Eskes1990}
\bibinfo{author}{\bibfnamefont{H.}~\bibnamefont{Eskes}},
  \bibinfo{author}{\bibfnamefont{L.~H.} \bibnamefont{Tjeng}}, \bibnamefont{and}
  \bibinfo{author}{\bibfnamefont{G.~A.} \bibnamefont{Sawatzky}},
  \bibinfo{journal}{Phys. Rev. B} \textbf{\bibinfo{volume}{41}},
  \bibinfo{pages}{288} (\bibinfo{year}{1990}),
  \urlprefix\url{https://link.aps.org/doi/10.1103/PhysRevB.41.288}.

\bibitem[{\citenamefont{Armitage et~al.}(2010)\citenamefont{Armitage, Fournier,
  and Greene}}]{Armitage2010}
\bibinfo{author}{\bibfnamefont{N.~P.} \bibnamefont{Armitage}},
  \bibinfo{author}{\bibfnamefont{P.}~\bibnamefont{Fournier}}, \bibnamefont{and}
  \bibinfo{author}{\bibfnamefont{R.~L.} \bibnamefont{Greene}},
  \bibinfo{journal}{Rev. Mod. Phys.} \textbf{\bibinfo{volume}{82}},
  \bibinfo{pages}{2421} (\bibinfo{year}{2010}),
  \urlprefix\url{https://link.aps.org/doi/10.1103/RevModPhys.82.2421}.

\bibitem[{\citenamefont{Haule et~al.}(2014)\citenamefont{Haule, Birol, and
  Kotliar}}]{Haule2014}
\bibinfo{author}{\bibfnamefont{K.}~\bibnamefont{Haule}},
  \bibinfo{author}{\bibfnamefont{T.}~\bibnamefont{Birol}}, \bibnamefont{and}
  \bibinfo{author}{\bibfnamefont{G.}~\bibnamefont{Kotliar}},
  \bibinfo{journal}{Phys. Rev. B} \textbf{\bibinfo{volume}{90}},
  \bibinfo{pages}{075136} (\bibinfo{year}{2014}),
  \urlprefix\url{https://link.aps.org/doi/10.1103/PhysRevB.90.075136}.

\bibitem[{\citenamefont{Chen et~al.}(2021{\natexlab{b}})\citenamefont{Chen,
  Wang, Rebec, Jia, Hashimoto, Lu, Moritz, Moore, Devereaux, and
  Shen}}]{Chen2021}
\bibinfo{author}{\bibfnamefont{Z.}~\bibnamefont{Chen}},
  \bibinfo{author}{\bibfnamefont{Y.}~\bibnamefont{Wang}},
  \bibinfo{author}{\bibfnamefont{S.~N.} \bibnamefont{Rebec}},
  \bibinfo{author}{\bibfnamefont{T.}~\bibnamefont{Jia}},
  \bibinfo{author}{\bibfnamefont{M.}~\bibnamefont{Hashimoto}},
  \bibinfo{author}{\bibfnamefont{D.}~\bibnamefont{Lu}},
  \bibinfo{author}{\bibfnamefont{B.}~\bibnamefont{Moritz}},
  \bibinfo{author}{\bibfnamefont{R.~G.} \bibnamefont{Moore}},
  \bibinfo{author}{\bibfnamefont{T.~P.} \bibnamefont{Devereaux}},
  \bibnamefont{and} \bibinfo{author}{\bibfnamefont{Z.-X.} \bibnamefont{Shen}},
  \bibinfo{journal}{Science}  (\bibinfo{year}{2021}{\natexlab{b}}).

\bibitem[{\citenamefont{Qu et~al.}(2022)\citenamefont{Qu, Chen, Jiang, Wang,
  and Li}}]{Qu2022}
\bibinfo{author}{\bibfnamefont{D.-W.} \bibnamefont{Qu}},
  \bibinfo{author}{\bibfnamefont{B.-B.} \bibnamefont{Chen}},
  \bibinfo{author}{\bibfnamefont{H.-C.} \bibnamefont{Jiang}},
  \bibinfo{author}{\bibfnamefont{Y.}~\bibnamefont{Wang}}, \bibnamefont{and}
  \bibinfo{author}{\bibfnamefont{W.}~\bibnamefont{Li}},
  \bibinfo{journal}{Commun Phys} \textbf{\bibinfo{volume}{5}},
  \bibinfo{pages}{257} (\bibinfo{year}{2022}).

\bibitem[{\citenamefont{Peng et~al.}(2023)\citenamefont{Peng, Wang, Wen, Lee,
  Devereaux, and Jiang}}]{Peng2023}
\bibinfo{author}{\bibfnamefont{C.}~\bibnamefont{Peng}},
  \bibinfo{author}{\bibfnamefont{Y.}~\bibnamefont{Wang}},
  \bibinfo{author}{\bibfnamefont{J.}~\bibnamefont{Wen}},
  \bibinfo{author}{\bibfnamefont{Y.~S.} \bibnamefont{Lee}},
  \bibinfo{author}{\bibfnamefont{T.~P.} \bibnamefont{Devereaux}},
  \bibnamefont{and} \bibinfo{author}{\bibfnamefont{H.-C.} \bibnamefont{Jiang}},
  \bibinfo{journal}{Phys. Rev. B} \textbf{\bibinfo{volume}{107}},
  \bibinfo{pages}{L201102} (\bibinfo{year}{2023}),
  \urlprefix\url{https://link.aps.org/doi/10.1103/PhysRevB.107.L201102}.

\bibitem[{\citenamefont{K\"uhner et~al.}(2000)\citenamefont{K\"uhner, White,
  and Monien}}]{Kuhner2000}
\bibinfo{author}{\bibfnamefont{T.~D.} \bibnamefont{K\"uhner}},
  \bibinfo{author}{\bibfnamefont{S.~R.} \bibnamefont{White}}, \bibnamefont{and}
  \bibinfo{author}{\bibfnamefont{H.}~\bibnamefont{Monien}},
  \bibinfo{journal}{Phys. Rev. B} \textbf{\bibinfo{volume}{61}},
  \bibinfo{pages}{12474} (\bibinfo{year}{2000}),
  \urlprefix\url{https://link.aps.org/doi/10.1103/PhysRevB.61.12474}.

\bibitem[{\citenamefont{White et~al.}(2002)\citenamefont{White, Affleck, and
  Scalapino}}]{White2002}
\bibinfo{author}{\bibfnamefont{S.~R.} \bibnamefont{White}},
  \bibinfo{author}{\bibfnamefont{I.}~\bibnamefont{Affleck}}, \bibnamefont{and}
  \bibinfo{author}{\bibfnamefont{D.~J.} \bibnamefont{Scalapino}},
  \bibinfo{journal}{Phys. Rev. B} \textbf{\bibinfo{volume}{65}},
  \bibinfo{pages}{165122} (\bibinfo{year}{2002}).

\bibitem[{\citenamefont{Calabrese and Cardy}(2004)}]{Calabrese2004}
\bibinfo{author}{\bibfnamefont{P.}~\bibnamefont{Calabrese}} \bibnamefont{and}
  \bibinfo{author}{\bibfnamefont{J.}~\bibnamefont{Cardy}}, \bibinfo{journal}{J.
  Stat. Mech. Theory Exp.} \textbf{\bibinfo{volume}{2004}}
  (\bibinfo{year}{2004}).

\bibitem[{\citenamefont{Fagotti and Calabrese}(2011)}]{Fagotti2011}
\bibinfo{author}{\bibfnamefont{M.}~\bibnamefont{Fagotti}} \bibnamefont{and}
  \bibinfo{author}{\bibfnamefont{P.}~\bibnamefont{Calabrese}},
  \bibinfo{journal}{J. Stat. Mech. Theory Exp.} \textbf{\bibinfo{volume}{2011}}
  (\bibinfo{year}{2011}).

\bibitem[{\citenamefont{Jiang et~al.}(2018)\citenamefont{Jiang, Weng, and
  Kivelson}}]{Jiang2018tJ}
\bibinfo{author}{\bibfnamefont{H.-C.} \bibnamefont{Jiang}},
  \bibinfo{author}{\bibfnamefont{Z.-Y.} \bibnamefont{Weng}}, \bibnamefont{and}
  \bibinfo{author}{\bibfnamefont{S.~A.} \bibnamefont{Kivelson}},
  \bibinfo{journal}{Phys. Rev. B} \textbf{\bibinfo{volume}{98}},
  \bibinfo{pages}{140505} (\bibinfo{year}{2018}),
  \urlprefix\url{https://link.aps.org/doi/10.1103/PhysRevB.98.140505}.

\bibitem[{\citenamefont{Jiang and Devereaux}(2019)}]{Jiang2019Hub}
\bibinfo{author}{\bibfnamefont{H.-C.} \bibnamefont{Jiang}} \bibnamefont{and}
  \bibinfo{author}{\bibfnamefont{T.~P.} \bibnamefont{Devereaux}},
  \bibinfo{journal}{Science} \textbf{\bibinfo{volume}{365}},
  \bibinfo{pages}{1424} (\bibinfo{year}{2019}), ISSN \bibinfo{issn}{0036-8075},
  \urlprefix\url{https://science.sciencemag.org/content/365/6460/1424}.

\bibitem[{\citenamefont{Jiang et~al.}(2020{\natexlab{a}})\citenamefont{Jiang,
  Zaanen, Devereaux, and Jiang}}]{Jiang2020prr}
\bibinfo{author}{\bibfnamefont{Y.-F.} \bibnamefont{Jiang}},
  \bibinfo{author}{\bibfnamefont{J.}~\bibnamefont{Zaanen}},
  \bibinfo{author}{\bibfnamefont{T.~P.} \bibnamefont{Devereaux}},
  \bibnamefont{and} \bibinfo{author}{\bibfnamefont{H.-C.} \bibnamefont{Jiang}},
  \bibinfo{journal}{Phys. Rev. Research} \textbf{\bibinfo{volume}{2}},
  \bibinfo{pages}{033073} (\bibinfo{year}{2020}{\natexlab{a}}),
  \urlprefix\url{https://link.aps.org/doi/10.1103/PhysRevResearch.2.033073}.

\bibitem[{\citenamefont{Chung et~al.}(2020)\citenamefont{Chung, Qin, Zhang,
  Schollw\"ock, and White}}]{Chung2020}
\bibinfo{author}{\bibfnamefont{C.-M.} \bibnamefont{Chung}},
  \bibinfo{author}{\bibfnamefont{M.}~\bibnamefont{Qin}},
  \bibinfo{author}{\bibfnamefont{S.}~\bibnamefont{Zhang}},
  \bibinfo{author}{\bibfnamefont{U.}~\bibnamefont{Schollw\"ock}},
  \bibnamefont{and} \bibinfo{author}{\bibfnamefont{S.~R.} \bibnamefont{White}}
  (\bibinfo{collaboration}{The Simons Collaboration on the Many-Electron
  Problem}), \bibinfo{journal}{Phys. Rev. B} \textbf{\bibinfo{volume}{102}},
  \bibinfo{pages}{041106} (\bibinfo{year}{2020}),
  \urlprefix\url{https://link.aps.org/doi/10.1103/PhysRevB.102.041106}.

\bibitem[{\citenamefont{Jiang et~al.}(2020{\natexlab{b}})\citenamefont{Jiang,
  Chen, and Weng}}]{Jiang2020prb}
\bibinfo{author}{\bibfnamefont{H.-C.} \bibnamefont{Jiang}},
  \bibinfo{author}{\bibfnamefont{S.}~\bibnamefont{Chen}}, \bibnamefont{and}
  \bibinfo{author}{\bibfnamefont{Z.-Y.} \bibnamefont{Weng}},
  \bibinfo{journal}{Phys. Rev. B} \textbf{\bibinfo{volume}{102}},
  \bibinfo{pages}{104512} (\bibinfo{year}{2020}{\natexlab{b}}),
  \urlprefix\url{https://link.aps.org/doi/10.1103/PhysRevB.102.104512}.

\bibitem[{\citenamefont{Gong et~al.}(2021)\citenamefont{Gong, Zhu, and
  Sheng}}]{Gong2021}
\bibinfo{author}{\bibfnamefont{S.}~\bibnamefont{Gong}},
  \bibinfo{author}{\bibfnamefont{W.}~\bibnamefont{Zhu}}, \bibnamefont{and}
  \bibinfo{author}{\bibfnamefont{D.~N.} \bibnamefont{Sheng}},
  \bibinfo{journal}{Phys. Rev. Lett.} \textbf{\bibinfo{volume}{127}},
  \bibinfo{pages}{097003} (\bibinfo{year}{2021}),
  \urlprefix\url{https://link.aps.org/doi/10.1103/PhysRevLett.127.097003}.

\bibitem[{\citenamefont{Zhang and Rice}(1988)}]{Zhang1988}
\bibinfo{author}{\bibfnamefont{F.~C.} \bibnamefont{Zhang}} \bibnamefont{and}
  \bibinfo{author}{\bibfnamefont{T.~M.} \bibnamefont{Rice}},
  \bibinfo{journal}{Phys. Rev. B} \textbf{\bibinfo{volume}{37}},
  \bibinfo{pages}{3759} (\bibinfo{year}{1988}),
  \urlprefix\url{https://link.aps.org/doi/10.1103/PhysRevB.37.3759}.

\bibitem[{\citenamefont{Lee et~al.}(2006)\citenamefont{Lee, Nagaosa, and
  Wen}}]{Lee2006}
\bibinfo{author}{\bibfnamefont{P.~A.} \bibnamefont{Lee}},
  \bibinfo{author}{\bibfnamefont{N.}~\bibnamefont{Nagaosa}}, \bibnamefont{and}
  \bibinfo{author}{\bibfnamefont{X.-G.} \bibnamefont{Wen}},
  \bibinfo{journal}{Rev. Mod. Phys.} \textbf{\bibinfo{volume}{78}},
  \bibinfo{pages}{17} (\bibinfo{year}{2006}),
  \urlprefix\url{https://link.aps.org/doi/10.1103/RevModPhys.78.17}.

\bibitem[{\citenamefont{Eskes et~al.}(1989)\citenamefont{Eskes, Sawatzky, and
  Feiner}}]{Eskes1989}
\bibinfo{author}{\bibfnamefont{H.}~\bibnamefont{Eskes}},
  \bibinfo{author}{\bibfnamefont{G.}~\bibnamefont{Sawatzky}}, \bibnamefont{and}
  \bibinfo{author}{\bibfnamefont{L.}~\bibnamefont{Feiner}},
  \bibinfo{journal}{Physica C: Superconductivity}
  \textbf{\bibinfo{volume}{160}}, \bibinfo{pages}{424} (\bibinfo{year}{1989}),
  ISSN \bibinfo{issn}{0921-4534},
  \urlprefix\url{https://www.sciencedirect.com/science/article/pii/0921453489904152}.

\end{thebibliography}

\end{document}